\documentclass[useAMS,usenatbib]{mn2e}
\usepackage{natbib}
\usepackage{graphicx}
\bibpunct{(}{)}{;}{a}{}{.}

\title[Optical/X-ray connection in 20 galaxy clusters]
{The optical/X-ray connection: ICM iron content and
galaxy optical luminosity in 20 galaxy clusters}

\author[T. F. Lagan\'{a} et al.]
  {T.~F.~Lagan\'{a}$^1$\thanks{E-mail: tflagana@astro.iag.usp.br},
  R. A. Dupke$^2$, L.~Sodr\'{e} Jr.$^1$, G.~B.~Lima Neto$^1$, 
  F. Durret$^3$
     \\
  $^1$Instituto de Astronomia, Geof\'{i}sica e C. Atmosf./USP, R. do Mat\~{a}o 1226, 
05508-090. S\~{a}o Paulo/SP, Brazil\\
  $^2$ University of Michigan, Ann Arbor, MI 48109-1090, USA\\
   $^3$ Institut d'Astrophysique de Paris, CNRS, UMR 7095, Universit\'e
  Pierre et Marie Curie, 98bis Bd Arago, 75014 Paris, France}
\date{Accepted 2008 November 21. Received in original form 2008 August 13}

\pagerange{\pageref{firstpage}--\pageref{lastpage}} \pubyear{2008}

\def\LaTeX{L\kern-.36em\raise.3ex\hbox{a}\kern-.15em
    T\kern-.1667em\lower.7ex\hbox{E}\kern-.125emX}

\begin{document}

\maketitle

\begin{abstract}
X-ray observations of galaxy clusters have shown that the intra-cluster
gas has iron abundances of about one third of the solar value. These
observations also show that part (if not all) of the intra-cluster gas
metals were produced within the member galaxies. We present a systematic
analysis of 20 galaxy clusters to explore the connection between the iron
mass and the total luminosity of early-type and late-type galaxies, and
 of the
brightest cluster galaxies (BCGs).  From our results, the intra-cluster
medium (ICM) iron mass seems to correlate better with the luminosity of
the BCGs than with that of the red and blue galaxy populations.  As the BCGs
cannot produce alone the observed amount of iron, we suggest that
ram-pressure plus tidal stripping act together to enhance, at the same
time, the BCG luminosities and the iron mass in the ICM.  Through the
analysis of the iron yield, we have also estimated that SN Ia are
responsible for more than 50\% of the total iron in the ICM.  This
result corroborates the fact that ram-pressure contributes to the gas
removal from galaxies to the inta-cluster medium, being very efficient
for clusters in the temperature range $2 < kT$~(keV)~$< 10$.

\end{abstract}


\section{Introduction}
\label{intro} The detection of the Fe line from the X-ray spectral
analysis of the intra-cluster medium
\citep[ICM;][]{Mitchell76,Serlemitsos77} indicates that it does not have
a primordial chemical composition but was enriched with material
processed in stars.  The relative importance of the mechanisms that
transport metal rich gas to the ICM is not well known.  The measurement
of heavy element abundances in the ICM can provide important clues on
the chemical evolution inside galaxy clusters.

Difficulties in determining the nature of metal enrichment in clusters
are enhanced by the variety of astrophysical processes and spatial
scales involved: yields from different supernova (SNe) types, the role
of galactic winds \citep[e.g.,][]{Dupke00}, ram-pressure
\citep[e.g.,][]{GG72} and tidal stripping \citep[e.g.,][]{TT72} as
mechanisms of metal transport from galaxies to the ICM, star formation
efficiency, and the influence of the environment, among other issues.

An important step in clarifying this problem was taken by
\citet{Arnaud92}.  These authors investigated the correlations between
some properties of the ICM, like the gas mass, with the optical
luminosity in clusters, finding that the gas mass correlates well with
the luminosity of E+S0 galaxies but not with the luminosity of
spirals. For a sample of 6 clusters with measured iron abundances and at
low redshift, they also found a good correlation between the iron mass
and the luminosity of red galaxies (E+S0).  They then concluded that
ellipticals and lenticulars are dominant in enriching the ICM,
indicating that the proto-galactic winds driven by type II SNe at the
early stages of cluster formation play a major role in the ICM metal
enrichment.  Another relevant enrichment mechanism is ram-pressure
stripping \citep[hereafter RPS,][]{GG72}.  Current observations indicate
that RPS is more common than previously believed, acting also in low
density environments such as poor groups and in the outskirts of
clusters \citep{Solanes01,Kantharia05,Levy07,Kantharia08}.  This is also
supported by recent numerical simulations
\citep[e.g.,][]{Vollmer05,Hester06,Bruggen08}. Ram-pressure stripped gas
is more enriched by SN Ia ejecta and can continuously provide an inflow
of iron to the ICM through the cluster history \citep{Renzini93}.

Although the ICM metal enrichment problem has been addressed in many studies 
\citep[among others]{TNH90,Mushotzky96,ML97,Dupke00},
there are good reasons to revisit it. Firstly, \citet{Arnaud92} used a limited sample 
(the only 6 available clusters with masses and luminosities in the literature). Secondly,
we are now in the era of precision abundances determination given the superior 
ability for spatially resolved spectroscopy of the instruments on board of XMM-Newton, 
{\it Chandra} and Suzaku. Finally, there are currently a number of studies, providing
ICM physical parameters from large cluster samples homogeneously determined 
\citep[e.g.,][]{Zhang07,M08}.

To readdress this question, we investigate in this work  the relation between the
metallicity (iron abundance) of the intra-cluster gas and some optical
properties of galaxy clusters, in search for clues on how metals got into the ICM. 

The outline of the paper is as follows: in Sect.~\ref{data_red} we describe the sources
of the data analyzed. In Sect.~\ref{sel} we present the photometric 
analysis, describing the separation between the red and blue populations and
the determination of their total luminosity. Influence of the Butcher-Oemler effect is also discussed in this section. 
In Sect.~\ref{Fe_proc} we discuss the 
dominant population in the iron enrichment by means of the correlation between 
the iron mass and the total luminosity of the different populations, the mechanism 
responsible for the metal transport from galaxies to the ICM and the role of SN II 
and SN Ia. Finally, we present our conclusions in Sect.~\ref{conc}.

\section{The Data}
\label{data_red}
We describe in this section the X-ray and optical data used in our
analysis.

\subsection{X-ray Data}
\label{Xdata}

\citet{M08} carried out an X-ray analysis 
of a homogeneous sample of 115 galaxy clusters in 
the redshift range $0.1 < z < 1.3$. The sample was assembled from publicly 
available {\it Chandra} data and selected in order to 
ensure cluster emission detection up to $r_{500}$ (the radius within which 
the cluster mean density exceeds the critical value by a factor of 500), 
allowing cluster properties to be directly measured up to this radius. 
We have selected a subsample of galaxy clusters that were analyzed by \citet{M08} 
and observed by the {\it Sloan Digital Sky Survey} 
\citep[SDSS,][]{York2000}. Out of 115 clusters 20 were selected (Table~\ref{Tab21cl}), 
and will be the object of the analysis presented in this paper.
All the X-ray data used here (metallicity, gas mass and temperature) 
are taken from \citet{M08} and are also presented in Table~\ref{Tab21cl}.

\begin{figure}
\centering
\includegraphics[width=0.35\textwidth,angle=90,clip=true]{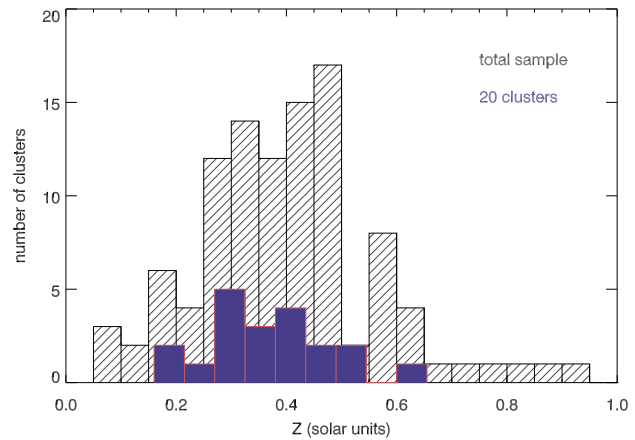}
\caption{\small Distribution of iron abundance determined within $r_{500}$ for our 
selected sub-sample (in dark blue) 
and for the whole sample analyzed by \citet{M08} (filled with dashed lines). 
Our mean value is $Z=(0.34  \pm 0.15) Z_{\odot}$, 
while the average mean for the whole sample is $Z=(0.39 \pm  0.14) Z_{\odot}$} 
\label{histoFe}
\end{figure}

In order to check if the 20 clusters analyzed in this work are a fair
sample of the original one, we present in Fig.~\ref{histoFe} the
metallicity (iron abundance) for all the clusters studied by
\citet{M08}.  The mean metallicity value ($Z=(0.34  \pm 0.15) Z_{\odot}$) obtained for our selected
sub-sample agrees well with the mean value ($Z=(0.39 \pm  0.14) Z_{\odot}$) 
obtained for the original 115 clusters.

\begin{table*}
\centering
\caption{\small Clusters from \citet{M08} data sample which were observed by
SDSS and have X-ray flux high enough for the analysis presented here. 
Column~(1): cluster name; Col.~(2): right ascension; Col.~(3):
declination; Col.~(4): redshift; Col.~(5): $r_{500}$ radius; Col.~(6): mean temperature 
derived from the X-ray emission within $r_{500}$; Col.~(7): mean
metallicity inside within $r_{500}$; Col.~(8): gas mass within $r_{500}$.}
\begin{tabular}{cccccccc}
\hline\hline
Cluster & R.A & DEC & {\it z} & $r_{500}$ & $<kT>$ & $<Z>$&   $M_{gas}$ \\
         & (J2000) & (J2000)& & ($h_{70}^{-1})$ Mpc &  (keV)& & $10^{13} M_{\odot}$ \\
\hline\noalign{\smallskip}
A267          & 01:52:42.12 & +01:00:41.4 & 0.230 &1.04& $4.9_{-0.3}^{+0.3}$ & $0.49_{-0.17}^{+0.18}$  & $5.74_{-0.07}^{+0.10}$\\
MS0906.5+1110 & 09:09:12.72 & +10:58:33.6 & 0.180 &1.06& $5.3_{-0.2}^{+0.2}$ & $0.31_{-0.07}^{+0.07}$  & $5.27_{-0.04}^{+0.05}$\\
A773          & 09:17:52.80 & +51:43:40.4 & 0.217 &1.25& $7.4_{-0.3}^{+0.3}$ & $0.48_{-0.06}^{+0.06}$  & $9.09_{-0.06}^{+0.06}$\\
MS1006.0+1202 & 10:08:47.52 & +11:47:40.6 & 0.221 &1.11& $5.9_{-0.4}^{+0.4}$ & $0.16_{-0.10}^{+0.10}$  & $5.29_{-0.06}^{+0.06}$ \\
A1204         & 11:13:20.40 & +17:35:39.1 & 0.171 &0.92& $3.4_{-0.1}^{+0.1}$ & $0.37_{-0.05}^{+0.05}$  & $3.10_{-0.06}^{+0.21}$ \\
A1240         & 11:23:37.68 & +43:05:44.5 & 0.159 &0.92& $3.9_{-0.3}^{+0.3}$ & $0.19_{-0.09}^{+0.10}$  & $2.75_{-0.03}^{+0.04}$ \\
A1413         & 11:55:18.00 & +23:24:17.6 & 0.143 &1.26& $7.2_{-0.2}^{+0.2}$ & $0.41_{-0.03}^{+0.03}$  & $7.95_{-0.05}^{+0.04}$ \\
A1682         & 13:06:51.12 & +46:33:29.5 & 0.234 &1.13& $6.2_{-0.8}^{+0.8}$ & $0.42_{-0.25}^{+0.27}$  & $7.07_{-0.12}^{+0.13}$ \\
A1689         & 13:11:29.52 & -01:20:30.4 & 0.183 &1.37& $9.0_{-0.3}^{+0.3}$ & $0.42_{-0.04}^{+0.04}$  & $11.26_{-0.06}^{+0.11}$ \\
A1763         & 13:35:18.24 & +40:59:59.3 & 0.223 &1.32& $7.8_{-0.4}^{+0.4}$ & $0.29_{-0.07}^{+0.07}$  & $11.47_{-0.07}^{+0.10}$ \\
A1914         & 14:26:01.20 & +37:49:35.4 & 0.171 &1.37& $9.8_{-0.3}^{+0.3}$ & $0.34_{-0.05}^{+0.05}$  & $10.69_{-0.07}^{+0.09}$ \\
A1942  	      & 14:38:22.08 & +03:40:06.2 & 0.224 &0.94& $4.3_{-0.2}^{+0.3}$ & $0.27_{-0.08}^{+0.08}$  & $3.59_{-0.03}^{+0.03}$ \\
RXJ1504-0248  & 15:04:07.44 & -02:48:18.4 & 0.215 &1.34& $6.8_{-0.2}^{+0.2}$ & $0.35_{-0.04}^{+0.04}$  & $10.9_{-0.47}^{+0.04}$ \\
A2034         & 15:10:12.48 & +33:30:28.4 & 0.113 &1.22& $6.7_{-0.2}^{+0.2}$ & $0.38_{-0.04}^{+0.04}$  & $6.88_{-0.03}^{+0.02}$ \\
A2069         & 15:24:09.36 & +29:53:10.0 & 0.116 &1.20& $6.3_{-0.2}^{+0.2}$ & $0.29_{-0.05}^{+0.05}$  & $6.56_{-0.03}^{+0.03}$\\
A2111         & 15:39:41.28 & +34:25:10.2 & 0.229 &1.18& $6.8_{-0.5}^{+0.9}$ & $0.23_{-0.15}^{+0.15}$  & $7.44_{-0.05}^{+0.10}$ \\
RXJ1701+6414  & 17:01:23.04 & +64:14:11.4 & 0.225 &0.93& $5.2_{-0.4}^{+0.6}$ & $0.60_{-0.17}^{+0.18}$  & $4.21_{-0.06}^{+0.028}$ \\
A2259         & 17:20:08.64 & +27:40:09.8 & 0.164 &1.09& $5.6_{-0.4}^{+0.4}$ & $0.28_{-0.10}^{+0.11}$  & $5.29_{-0.07}^{+0.08}$ \\
RXJ1720.1+2638& 17:20:10.08 & +26:37:29.3 & 0.164 &1.24& $6.1_{-0.1}^{+0.1}$ & $0.48_{-0.03}^{+0.03}$  & $7.46_{-0.01}^{+0.31}$ \\
RXJ2129.6+0005& 21:29:40.08 & +00:05:19.6 & 0.235 &1.20& $5.6_{-0.3}^{+0.3}$ & $0.50_{-0.10}^{+0.10}$  & $8.02_{-0.14}^{+0.14}$ \\
\hline	  
\end{tabular}
\\
\label{Tab21cl}
\end{table*}

\subsection{SDSS data}

The  photometric data used in our analysis comes from SDSS Data Release 5 
\citep[DR5,][]{refDR5}, which
contains spectroscopic and photometric data for a large number of 
galaxies observed within almost a quarter of the sky. 

In order to investigate the correlation between X-ray properties and galaxy 
populations, we have selected from the SDSS data 
galaxies brighter than $r=22$ (extracted from GALAXY tables, which has primary objects subset with 
type ``galaxy'') and 
within the $r_{500}$ radius of each cluster centre \citep[adopted as the
X-ray centre from][]{M08}. This magnitude limit ensures a significant number 
of objects for the galaxy population analysis. In this study we will
adopt the $(g-r)$ colour, from the DERED tables which are tables of magnitudes
already corrected for galactic extinction.

We have also downloaded the same type of data in an annular control field 
around each cluster (see Section~\ref{sel}), for the statistical 
subtraction of foreground/background galaxies. 

\section{Photometric analysis}
\label{sel}

Since the number of galaxies with measured redshifts in the field of each cluster is small, we
have adopted a statistical approach, based on the analysis of the
colour-magnitude diagram (CMD) of each cluster, to estimate some properties of 
the cluster galaxies.

\begin{figure}
\includegraphics[width=60mm,angle=90]{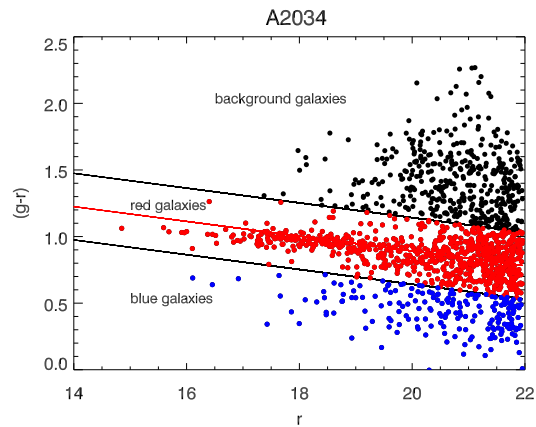}
\caption{Color-magnitude diagram for galaxies inside $r_{500}$ for A2034, divided in the three groups
discussed in the text: red, blue and background galaxies.}
\label{pop}
\end{figure}

We present in Fig.~\ref{pop} the $(g-r)$ versus $r$~ CMD of A2034 for all 
galaxies within $r_{500}$. 
Early-type galaxies in clusters present a tight correlation between their 
optical colours and luminosity, known as colour-magnitude relation or
red-sequence (RS), which is very useful for their identification. In general,
galaxies significantly above the red sequence are in the cluster background (since they are
redder than the reddest cluster galaxies). We call blue galaxies those 
below the red sequence (see Fig.~\ref{pop}), which comprise 
clusters and non-clusters members. 

We have used the control fields to  obtain the field contamination and 
the correction required to estimate the numbers
and luminosities of cluster galaxies.
The procedure we adopt is the following: the first step is to determine, for
each field, the position of the red sequence. 
\citet{Bower92}, \citet{Gladders98} and \citet{Romeo08}, among others, 
have shown that the slope of the red sequence
in CMDs is almost constant up to redshift $z=0.5$. For this reason, given
that all clusters in our sample have redshifts well below this value, we
assume that this slope is the same for all clusters in Table~\ref{Tab21cl}. 
Thus, we have used the CMDs of the 5 clusters of our sample (MS0906.5+1110, A2034, A2069, A2259 and RXJ1720.1+2638) 
with the most prominent red sequence to estimate this slope through a linear fit, 
obtaining $\sim -0.055$ (see Fig.~\ref{pop}).

For all clusters, we fitted the red sequence as a function of r-band magnitude
according to
\begin{equation}
\label{eqcor}
(g-r) =-0.055~r + A,
\end{equation}
to obtain the zero-point coefficient ($A$).
Then, in each cluster, we divided the galaxies in three groups:
\begin{itemize}
\item{{\it red galaxies}: those galaxies within 0.3 mag \citep{DLP08} from the best fit color-magnitude relation.
That is $(g-r) \pm 0.3$ mag zone.}
\item{{\it blue galaxies}: galaxies bluer than the $(g-r) - 0.3$ mag. 
This group contains mainly star-forming galaxies, both in the cluster and 
in the field.}
\item{{\it background galaxies}: galaxies redward of the upper limit of the 
red sequence. We assume that these galaxies are not cluster members.}
\end{itemize}

\begin{figure}
\centering
\includegraphics[width=0.30\textwidth,angle=90]{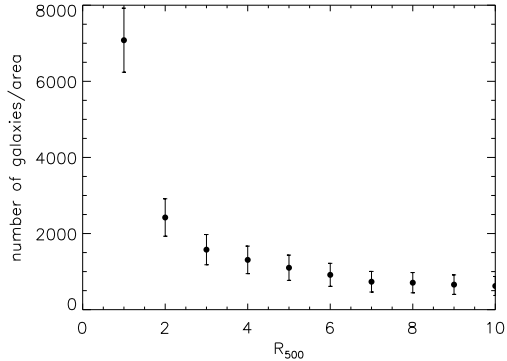}
\caption{Galaxy counts distribution as a function of the mean radius of 
each annulus, given as a function of $r_{500}$, for A1689. 
The background is estimated from the annuli with $R \ge 7 \times r_{500}$.}
\label{bkg_cts}
\end{figure}

\begin{figure*}
\centering
\includegraphics[width=0.7\textwidth,angle=90]{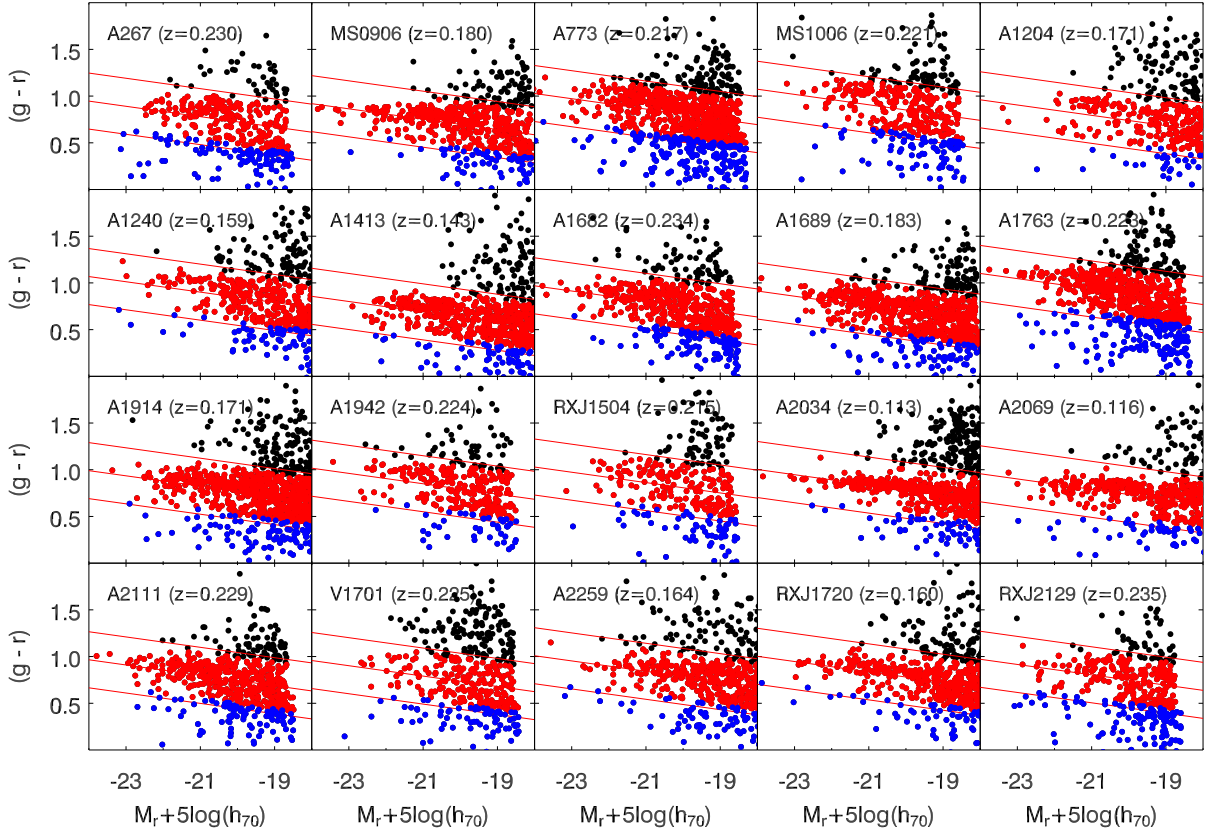}
\caption{Colour-magnitude diagrams within $r_{500}$ for the 20 clusters in our sample.}
\label{sv}
\end{figure*}

We estimate the cluster contamination by background/foreground galaxies
using a control field around the cluster defined as
an annular region between $7 \times r_{500}$ and $8 \times r_{500}$
from the cluster centre. This region is far enough from the centre
to allow the estimation of background densities, 
as shown in Fig.~\ref{bkg_cts}. 
We used the CMDs of the galaxies in the control fields to estimate the 
background contamination (in the number of galaxies and luminosity) for
red and blue galaxies in the cluster. The CMDs for the clusters in our 
sample are shown in Fig.~\ref{sv}.

For each bin of magnitude, the number count of galaxies is given by:
\begin{equation}
N_{\rm cl}(m)=N_{\rm cl+bkg}(m)-\gamma N_{\rm bkg}(m),
\end{equation}
where $N_{\rm cl}(m)$ is the expected number of cluster galaxies at a
certain magnitude interval, $N_{\rm cl+bkg}(m)$ is the total number of
galaxies along the line of sight in the same magnitude interval, $N_{\rm
bkg}(m)$ is the number of background and foreground galaxies in the same
magnitude interval, $\gamma$ is the ratio between cluster and control
field areas, and $m$ is the mean magnitude of the interval.  After
subtracting the background and foreground contamination, cluster counts
per magnitude bin in the $r$ band were determined and converted to
luminosity.

\subsection{Luminosity of the brightest galaxies}
The division of our sample in groups described above enable us to calculate 
separately the total luminosity of the red and the blue population. 
We obtained the $r$-band luminosity of the brightest cluster galaxies (BCGs), $L_{BCG}$, 
directly from the SDSS.

We used the distance modulus given by:
\begin{equation}
M_r = m_r - 25 - 5 \log(d_{\rm L}/ \rm 1 Mpc) - k(z),
\end{equation}
where $d_{L}$ is the luminosity distance (computed assuming $\Omega_m=0.3$, $\Omega_\lambda=0.7$,
and $h=0.7$) and $k$(z) is the $k$-correction computed by \citet{Poggianti97}.
Converting magnitudes to luminosities we integrated the luminosity function in order to
obtain the total luminosity for each population. These luminosities are given in Table~\ref{Tab_Ltot}.
\begin{table*}
\centering
\caption{Properties derived from optical data. Column (1): cluster name;  Col.(2): total 
luminosity of the red population within $r_{500}$; Col.(3): total luminosity of 
the blue population  within $r_{500}$; Col.(4): luminosity of BCGs; Col.(5): iron 
mass derived from Eq.~\ref{calc_Mfe} within $r_{500}$.}
\begin{tabular}{ccccc}
\hline\hline
Cluster &  $L_{red}$  & $L_{blue}$ &  $L_{BCG}$ & $M_{Fe}$ \\
& ($10^{12} L_{\odot}$) & ($10^{12} L_{\odot}$)  & ($10^{11} L_{\odot}$)& ($10^{11} M_{\odot}$) \\
\hline
A267		&	2.19	$\pm$	0.03	&	1.82	$\pm$	0.01	&	1.84	&	0.52	$\pm$	0.18	\\
MS0906.5+1110	&	2.69	$\pm$	0.05	&	1.00	$\pm$	0.01	&	0.92	&	0.30	$\pm$	0.06	\\
A773		&	3.82	$\pm$	0.07	&	2.08	$\pm$	0.01	&	0.68	&	0.80	$\pm$	0.10	\\
MS1006.0+1202	&	1.85	$\pm$	0.05	&	1.45	$\pm$	0.01	&	1.22	&	0.16	$\pm$	0.09	\\
A1204		&	0.66	$\pm$	0.02	&	0.74	$\pm$	0.02	&	0.54	&	0.21	$\pm$	0.03	\\
A1240		&	0.79	$\pm$	0.02	&	0.73	$\pm$	0.07	&	0.13	&	0.10	$\pm$	0.05	\\
A1413		&	0.79	$\pm$	0.04	&	0.83	$\pm$	0.01	&	1.71	&	0.60	$\pm$	0.04	\\
A1682		&	2.9	$\pm$	0.08	&	2.34	$\pm$	0.02	&	0.94	&	0.54	$\pm$	0.32	\\
A1689		&	2.31	$\pm$	0.04	&	1.21	$\pm$	0.07	&	0.76	&	0.87	$\pm$	0.01	\\
A1763		&	3.21	$\pm$	0.02	&	2.18	$\pm$	0.03	&	0.83	&	0.61	$\pm$	0.02	\\
A1914		&	1.77	$\pm$	0.03	&	0.92	$\pm$	0.04	&	0.46	&	0.67	$\pm$	0.09	\\
A1942		&	2.22	$\pm$	0.08	&	1.44	$\pm$	0.01	&	0.89	&	0.18	$\pm$	0.05	\\
RXJ1504-0248	&	1.66	$\pm$	0.04	&	1.35	$\pm$	0.08	&	1.66	&	0.70	$\pm$	0.05	\\
A2034		&	0.53	$\pm$	0.02	&	0.35	$\pm$	0.06	&	0.87	&	0.48	$\pm$	0.05	\\
A2069		&	0.89	$\pm$	0.04	&	0.34	$\pm$	0.04	&	0.41	&	0.35	$\pm$	0.06	\\
A2111		&	4.16	$\pm$	0.02	&	2.06	$\pm$	0.02	&	0.54	&	0.31	$\pm$	0.20	\\
RXJ1701+6414	&	2.02	$\pm$	0.05	&	1.59	$\pm$	0.01	&	0.06	&	0.46	$\pm$	0.16	\\
A2259		&	0.92	$\pm$	0.05	&	1.02	$\pm$	0.07	&	1.22	&	0.27	$\pm$	0.09	\\
RXJ1720.1+2638	&	1.13	$\pm$	0.03	&	0.94	$\pm$	0.07	&	1.23	&	0.66	$\pm$	0.38	\\
RXJ2129.6+0005	&	2.15	$\pm$	0.01	&	2.08	$\pm$	0.02	&	1.15	&	0.73	$\pm$	0.14	\\
\hline
\end{tabular}
\label{Tab_Ltot}
\end{table*}

\subsection{Butcher-Oemler effect}
In a hierarchical scenario clusters grow through the accretion of galaxies and groups,
most of them containing blue galaxies. Indeed, many blue galaxies in clusters may
be considered newcomers \citep[e.g.,][]{Sodre89} and, 
through  effects suffered in the hostile environment of
galaxy clusters, most of their ISM gas ends up being transferred to the ICM,
contributing to its chemical enrichment.
We call attention to the fact that the time-scale of colour 
(spectral) evolution is shorter than that of the morphological evolution \citep{Goto04}, 
meaning that 
some passive spiral galaxies through their evolution may be observed as red objects.

\begin{figure}
\centering
\includegraphics[width=0.35\textwidth,angle=90]{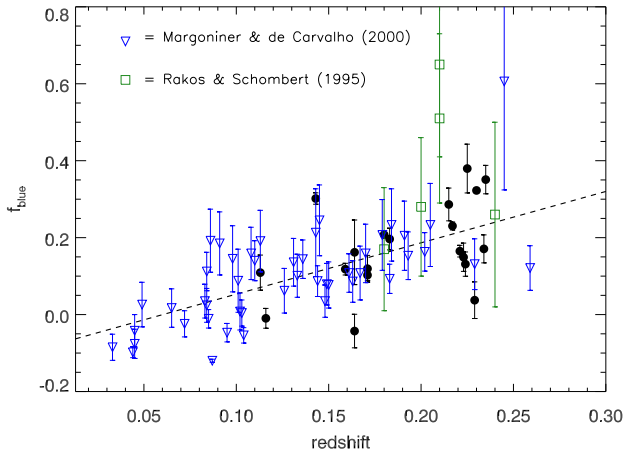}
\caption{\small Spiral fraction as a function of redshift. 
The dashed line is our best-fit for this sample. The filled circles are the sample analyzed i
n this work, while the green points were taken from \citet{RS95} and the upside down blue triangle 
are taken from \citet{Margo00} for this redshift interval.}
\label{BO}
\end{figure}

An observational evidence of the hierarchical scenario is the 
Butcher-Oemler (BO) effect \citep{BO84}. 
We have noticed that this effect can be detected
with our data \citep[and has been previously detected by][]{RS95,Margo00} and Figure~\ref{BO} 
shows that this trend can be detected even in low redshift clusters.
We define the fraction of blue galaxies as the ratio between the number 
of blue (see Sect.~\ref{sel}) to total number of galaxies in a cluster after background 
correction, indicating an extremely rapid change in the fraction of blue galaxies from  
$\sim 10 \%$ to more than $30 \%$ at only $z \sim 0.25$. 
In contrast, evolutionary models for passive evolution of simple stellar populations
predict that colour changes were not as dramatic as found here \citep{RS95}.

\section{The iron content/optical light connection in 20 galaxy clusters}
\label{Fe_proc}

In this section we investigate the links between the iron content and the luminosities 
of the red and blue populations of each cluster.

\subsection{The iron mass and its relation to the luminosity of the galaxy populations within $r_{500}$}

In clusters, there is a number of emission lines 
over the continuum bremsstrahlung spectrum, 
the most prominent being those in the iron complex at $\sim$ 7 keV. 
X-ray observations, have revealed the presence of 
heavy elements in the ICM, providing direct evidence that this gas was enriched by metals 
processed by stars \citep[e.g.,][]{Mitchell76,DFJ91,Mushotzky96}. 
Studies of large samples of clusters have shown that the Fe abundance is distributed around 
a value of 0.3 solar \citep{ML97,Allen98,Tozzi03,Ettori05}, which is consistent with the
mean value of our sample, $Z=0.34 \pm 0.15$.

The metallicity derived from X-ray spectra is emission-weighted, 
so that the values from the central regions tend to overpower global spectral fittings.
Therefore, a single global abundance value may lead to erroneous results when 
associating metals to galaxy luminosities in the whole cluster.
Chandra's excellent spatial resolution is a greatly advantageous over
previous satellites for this analysis
for intermediate redshift clusters because it can separate the inner and
outer regions extremely
well (not including projection effects), so that accurate average
abundances out of the cooling core region can be measured, 
making our analysis more realistic. While \citet{Arnaud92} used one
single value for abundances, we are using two different values (one for
the internal regions, within $0.15~r_{500}$,
and one for the annulus $0.15~r_{500} < r < r_{500}$), 
and the results are very similar (section 4.2). 
Naturally, one single value for the outer abundances is still susceptible 
to the above mentioned biases, but there were
simply not enough photons to subdivide the outer regions in several
annuli and still have abundances
with satisfactory significance.

In the present work, the iron mass of each cluster enclosed within $r_{500}$ 
was estimated as the product of the iron abundance by the gas mass and by the solar 
photospheric abundance by mass \citep[0.00197;][]{AG89}:
\begin{equation}
\label{calc_Mfe}
M_{Fe}=M_{\rm gas}(< r_{500})~\times~Z~\times~0.00197,
\end{equation}
where $M_{\rm gas}$ is the gas mass and $Z$ is the metallicity, both computed within $r_{500}$.

Eq.(\ref{calc_Mfe}) gives the relation between the iron mass in the ICM and the 
gas mass for our sample. The iron mass is consistent with a linear scaling 
with the gas mass, which is expected if $Z_{Fe}\approx$ constant.

\begin{figure*}
\centering
\includegraphics[width=0.45\textwidth,angle=90]{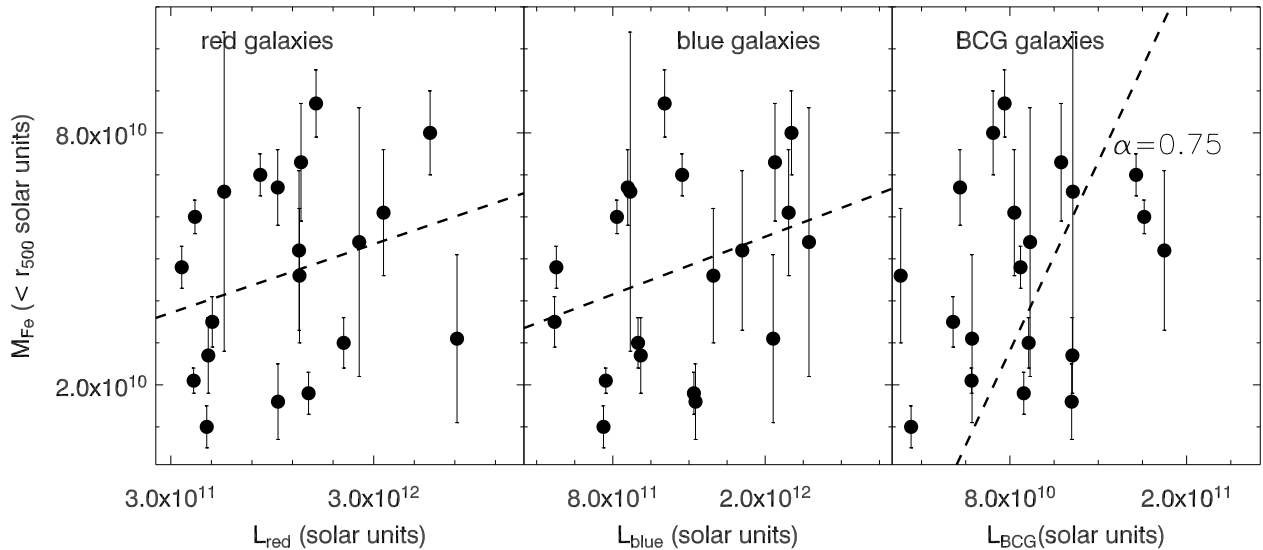}
\caption{\small Left panel: iron content as a function of total luminosity for
red galaxies. Middle panel: iron content as a function of total luminosity for
blue galaxies. Right panel: iron content as a function of BCG luminosities. 
The regression line has a slope
of $\alpha =0.75$. The 1$\sigma$ error on $\alpha$ is $\pm$ 0.33}
\label{FeL}
\end{figure*}

To explore what is the dominant population responsible for the metal enrichment in 
galaxy clusters we show, in Fig.~\ref{FeL}, the dependence of the iron mass
with the luminosity of red, blue and BCG galaxies. Given the large scatter observed, we adopt the 
robust Spearman correlation coefficient $\rho$ and the null probability (NP; i.e, the
probability of failing to reject the null hypothesis)  to 
evaluate the significance of the correlations \citep[see][]
{Press92}. 
We obtain $\rho$ values equal to 0.021 (NP=92\%), 0.15 (NP=52\%) and 0.41 (NP=8\%) for the red, 
blue and BCG galaxies,
respectively. 
The statistical significance of the correlations between luminosities
of different populations and ICM iron mass is very low, as seen in Fig.~\ref{FeL}.
Figures \ref{FeL} show that all the correlations are very poor and it is
not obvious that a given galaxy population is more significantly
correlated with the iron mass than the others. 
To address this question, we adopted a bootstrap resampling
technique to estimate for what fraction of a sample a correlation is
better than another in the sense that it has a larger Spearman
rank-order coefficient $\rho$ (Press et al. 1992).
We found that the correlation between the iron mass
and BCG luminosities is better than that with the luminosities of the
red galaxies at the 90\% confidence level and of the blue population at
the 85\% confidence level.
We also found that the significance
of the correlations between the blue and red populations with iron mass
are barely distinguishable and, in consequence, we cannot draw any
firm conclusion about the relative importance of these two populations in the ICM
iron enrichment. 

Our results do not confirm the classic idea where  
early-type galaxies play the major role in the ICM enrichment \citep{Arnaud92}. 
\citet{Arnaud92} included the BCGs in their `E+S0' sample and, from our work,
we find that when they are excluded from the red population, the correlation between the 
red galaxies and the iron mass becomes significantly worse (indistinguishable from
the blue galaxies). The most prominent correlation is that between the iron mass and the BCGs luminosities.
Note that their best fit for the iron mass as a function of the `E+S0' luminosities has a slope
of $\alpha=1.0 \pm 0.25$, which is, given the uncertainties, consistent with the poor
correlation obtained here between the BCGs and the iron mass. Including the
BCGs in the sample of red galaxies changes the correlation coefficient between the total galaxy
luminosity and the iron mass from  $\rho$=0.021 to $\rho$=0.31, closer to that found for BCGs alone,
as expected for the case of contamination by the BCGs.

\begin{figure*}
\centering
\includegraphics[width=0.45\textwidth,angle=90]{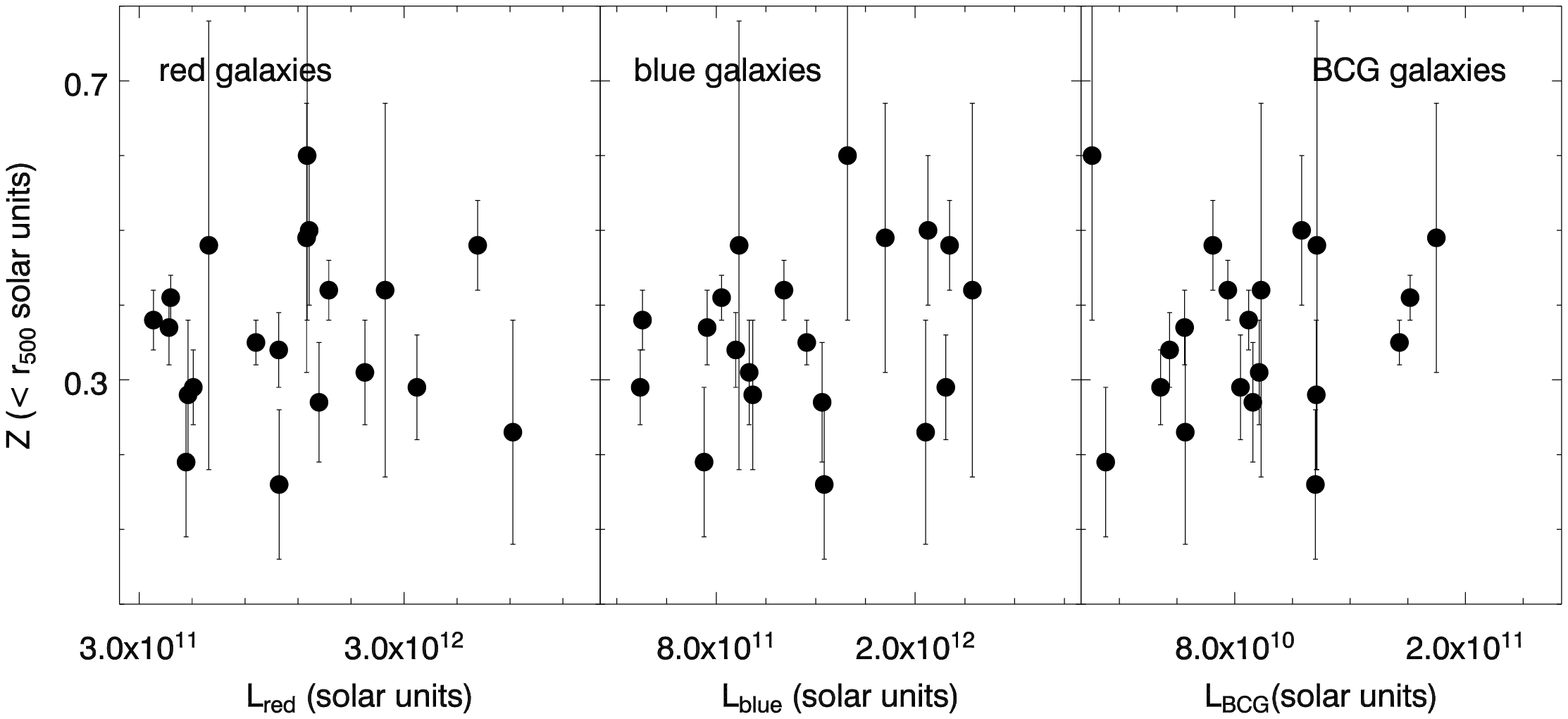}
\includegraphics[width=0.45\textwidth,angle=90]{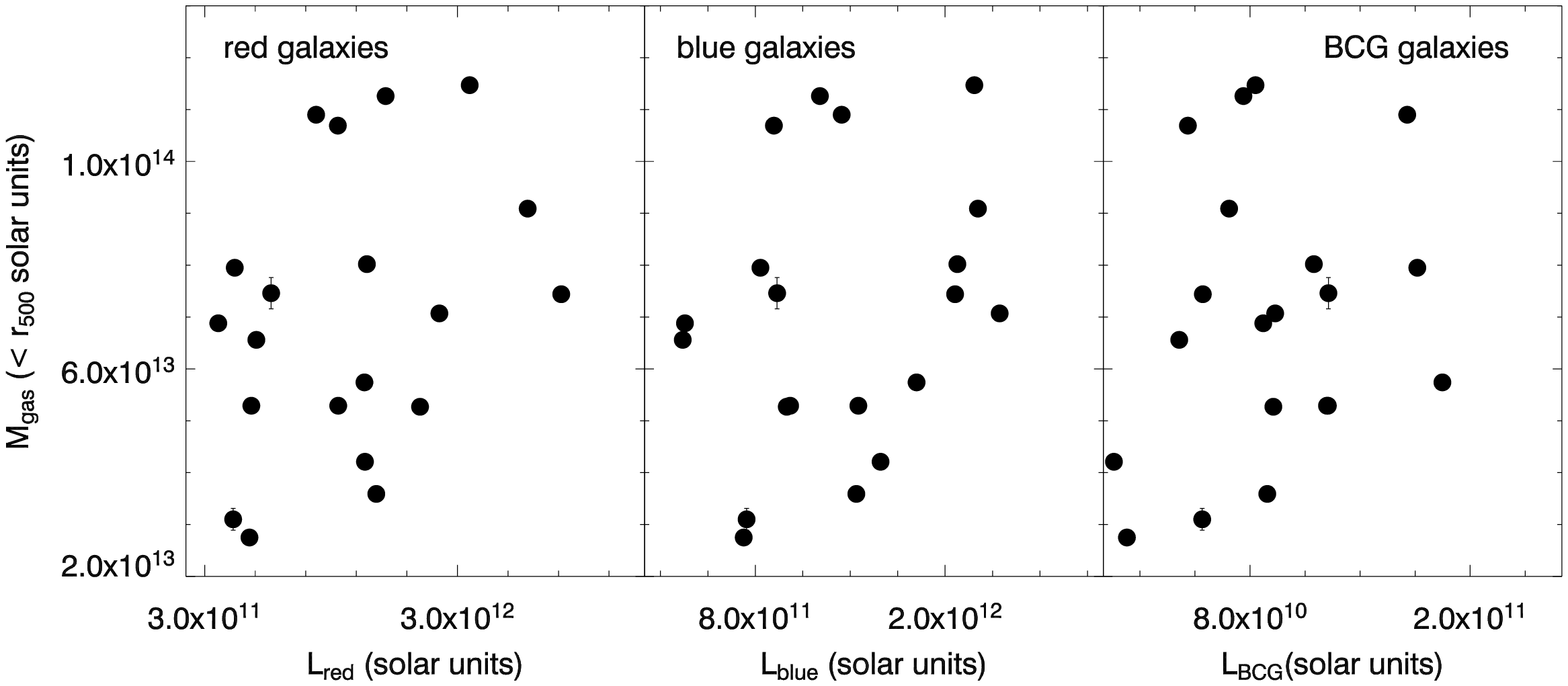}
\caption{\small Upper panels: Fe abundance within $r_{500}$ as a function of the total luminosity of the 
red galaxies (left panel), total luminosity of the blue galaxies (middle panel) 
and as a function of the BCG luminosities (right panel). Botton panels: gas mass as a functions of 
the total luminosity of the red galaxies (left panel), total luminosity of the blue galaxies (miidle panel) 
and as a function of the BCG luminosities. We do not found any significant trend in these 
panels}
\label{Lums}
\end{figure*}

Since the iron mass scales with the gas mass
and with the Fe abundance, we plot separately the relations
between the metallicities (see upper panels of Fig.~\ref{Lums})
and gas masses (see botton panels of Fig.~\ref{Lums}) versus the optical 
galaxy luminosities. From these figures, we still have low statistical corelations but
the ones between the BCGs luminosities and either the gas mass or the metallicity are
somewhat more visible than the others. We obtained correlation coefficients equal to $\rho$=0.075, 
$\rho$=0.17 and $\rho$=0.26 for the metallicity and the luminosities of the red, blue and BCGs galaxies, respectively. 
For the correlations between the gas mass and the luminosities, we obtained $\rho$=0.31, $\rho$=0.21 and $\rho$=0.33
for the red, blue and BCG population, respectively. The slightly better
correlation of gas mass with $L_{red}$ (as opposed to $L_{blue}$) is consistent
with the enhancement of the fraction of elliptical galaxies with
richness \citep{Dressler80}, given the somewhat high temperature range of
our sample (T $>$ 3.4 keV).

\subsection{The iron mass and its relation with the luminosity of 
the galaxy populations within an external annulus}

\citet{DeGrandi04} found that the iron mass associated to abundance excess which is normally found at the
center of cool-core clusters can be entirely produced by the BCGs. 
In order to take into account possible biases due to contamination by cool core processes, we
present here the same correlations computed within an external 
annulus, (0.15-1)$r_{500}$, to exclude the inner parts. 

The iron mass of each cluster enclosed within $(0.15 < r < 1) R_{500}$, 
was estimated from Eq.~(\ref{calc_Mfe}) and the gas mass was computed as,

\begin{equation}
M_{\rm gas}=\int_{0.15 r_{500}}^{r_{500}} \kern-1.5em \rho_{g} 4 \pi {r^{\prime}}^{2} \mathrm{d} r^{\prime} = 4 \pi m_{H} \overline{Z} \int_{0.15 r_{500}}^{r_{500}} \kern-1.5em {n_{e}(r^{\prime}) r^{\prime}}^{2} \mathrm{d} r^{\prime},
\end{equation}
where $\overline{Z}=1.25$ is the mean atomic weight for a H + He plasma, $m_{H}$ is the hydrogen mass and $n_{e}(r^{\prime})$ is the electron 
density which was fitted with a modified 
version of the standard $\beta$-model \citep[as proposed by][with $\gamma=3$ fixed for all fits]{Vikh06}. The best fit parameters were taken 
from Maughan (private communication). In Table~\ref{Tab_Ltot_anel} we present the values for the iron mass for our sample.

\begin{figure*}
\centering
\includegraphics[width=0.45\textwidth,angle=90]{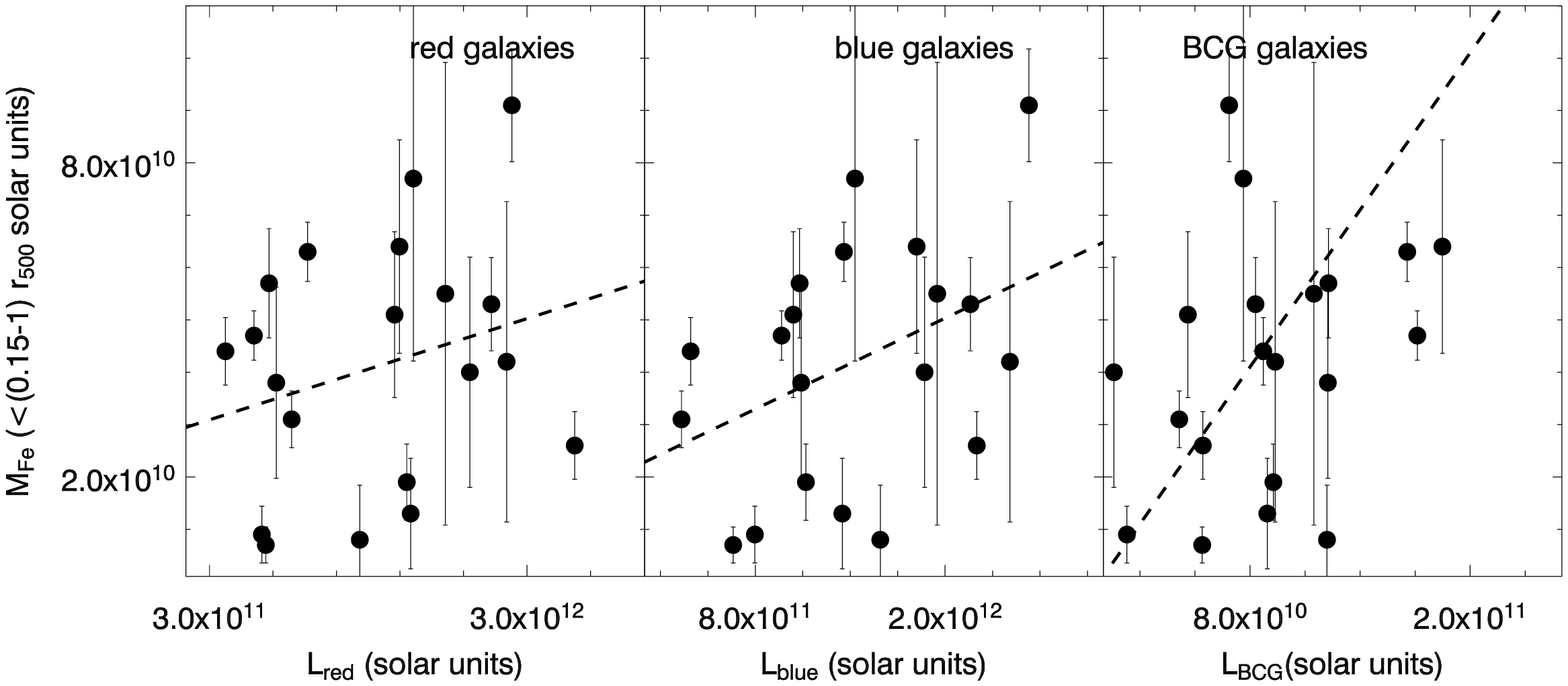}
\caption{\small Correlations between the total luminosities and the iron mass within (0.15-1)$r_{500}$.
Left panel: iron content as a function of the total luminosity of the 
red galaxies. Middle panel: iron content as a function of the total luminosity of the 
blue galaxies. Right panel: iron content as a function of the BCG luminosities.}
\label{FeL_anel}
\end{figure*}

We show in Fig.~\ref{FeL_anel} the plots between the iron mass and the luminosity of red, 
blue and BCG galaxies within (0.15-1) $r_{500}$.
We obtained $\rho$ values equal to 0.15 
(NP=48\%), 0.16 (NP=49\%) and 0.41 (NP=17\%) for the correlation between the iron mass and red, 
blue and BCG galaxies, respectively. Hence, we verified that even excluding the central region, where we
could find an iron excess due to the BCGs \citep{DeGrandi04}, the iron mass present in the ICM 
seems to still correlate better with the BCG population.  
In the next section we discuss a possible scenario to explain this dependence.

\begin{table*}
\centering
\caption{Properties derived from optical data. Column (1): cluster name;  Col.(2): total 
luminosity of the red population inside $(0.15-1)\ r_{500}$; Col.(3): total luminosity of 
the blue population inside $(0.15-1) r_{500}$; Col.(4): luminosity of BCGs; Col.(5): iron 
mass derived from Eq.~\ref{calc_Mfe} within $(0.15-1) r_{500}$.}
\begin{tabular}{cccccc}
\hline\hline
& \multicolumn{5}{c}{$(0.15 < r < 1) r_{500}$} \\
\cline{2-6} 
Cluster &  $L_{\rm red}$  & $L_{\rm blue}$ &   $<Z>$&   $M_{\rm gas}$ & $M_{\rm Fe}$ \\
& ($10^{12} L_{\odot}$) & ($10^{12} L_{\odot}$)  & & $10^{13} M_{\odot}$& ($10^{11} M_{\odot}$) \\
\hline
A267          &  2.10    $\pm$   0.02 & 1.82	$\pm$	0.12	&$0.64_{-0.24}^{+0.27}$ & 5.48 & 0.64$\pm$0.27\\
MS0906.5+1110 &  2.07	$\pm$	0.02 & 1.12	$\pm$	0.11	&$0.22_{-0.09}^{+0.09}$ & 4.67 & 0.19$\pm$0.07\\
A773          &  2.94	$\pm$	0.04 & 2.53	$\pm$	0.14	&$0.57_{-0.08}^{+0.09}$ & 8.72 & 0.91$\pm$0.14\\
MS1006.0+1202 &  1.70	$\pm$	0.05 & 1.59	$\pm$	0.15	&$0.09_{-0.09}^{+0.14}$ & 5.09 & 0.08$\pm$0.09\\
A1204         &  0.83	$\pm$	0.02 & 0.66	$\pm$	0.06	&$0.13_{-0.10}^{+0.10}$ & 2.81 & 0.07$\pm$0.02\\
A1240         &  0.76	$\pm$	0.02 & 0.80	$\pm$	0.08	&$0.18_{-0.10}^{+0.11}$ & 2.74 & 0.09$\pm$0.05\\
A1413         &  0.85	$\pm$	0.04 & 0.97	$\pm$	0.15	&$0.34_{-0.05}^{+0.05}$ & 7.50 & 0.47$\pm$0.07\\
A1682         &  2.92	$\pm$	0.06 & 2.41	$\pm$	0.23	&$0.33_{-0.30}^{+0.34}$ & 6.88 & 0.42$\pm$0.38\\
A1689         &  2.11	$\pm$	0.03 & 1.43	$\pm$	0.09	&$0.40_{-0.07}^{+0.07}$ & 10.48 & 0.77$\pm$0.02\\
A1763         &  2.78	$\pm$	0.12 & 2.16	$\pm$	0.30	&$0.26_{-0.09}^{+0.09}$ & 11.09 & 0.53$\pm$0.02\\
A1914         &  1.92	$\pm$	0.02 & 1.04	$\pm$	0.05	&$0.28_{-0.09}^{+0.09}$ & 9.97 & 0.51$\pm$0.16\\
A1942  	      &  2.10	$\pm$	0.05 & 1.35	$\pm$	0.13	&$0.20_{-0.08}^{+0.09}$ & 3.52 & 0.13$\pm$0.06\\
RXJ1504-0248  &  1.30	$\pm$	0.03 & 1.36	$\pm$	0.08	&$0.35_{-0.13}^{+0.14}$ & 9.88 & 0.63$\pm$0.25\\
A2034         &  0.52	$\pm$	0.02 & 0.39	$\pm$	0.07	&$0.36_{-0.05}^{+0.05}$ & 6.65 & 0.44$\pm$0.06\\
A2069         &  1.04	$\pm$	0.02 & 0.33	$\pm$	0.04	&$0.26_{-0.05}^{+0.06}$ & 6.44 & 0.31$\pm$0.06\\
A2111         &  3.46	$\pm$	0.09 & 2.20	$\pm$	0.24	&$0.20_{-0.18}^{+0.19}$ & 7.18 & 0.26$\pm$0.24\\
RXJ1701+6414  &  2.52	$\pm$	0.05 & 1.87	$\pm$	0.15	&$0.54_{-0.23}^{+0.26}$ & 4.04 & 0.40$\pm$0.19\\
A2259         &  0.99	$\pm$	0.05 & 1.09	$\pm$	0.08	&$0.41_{-0.16}^{+0.17}$ & 5.04 & 0.38$\pm$0.15\\
RXJ1720.1+2638&  0.93	$\pm$	0.02 & 1.08	$\pm$	0.08	&$0.45_{-0.07}^{+0.07}$ & 6.91 & 0.57$\pm$0.09\\
RXJ2129.6+0005&  2.42	$\pm$	0.09 & 1.95	$\pm$	0.14	&$0.40_{-0.18}^{+0.19}$ & 7.54 & 0.55$\pm$0.26\\
\hline	  	
\end{tabular}
\label{Tab_Ltot_anel}
\end{table*}

\subsection{The source of ICM metals}

The plots obtained in Figs.~\ref{FeL} and \ref{FeL_anel} suggest that
the BCGs could play a significant
role in the ICM iron enrichment. The most parsimonious scenario is that
a mechanism simultaneously contributes to enhance the metal mass and the luminosity of the BCG.
Such a mechanism can be, for example, RPS with tidal disruption of the galaxy near the center of the
cluster \citep[e.g.,][]{Cypriano06,Murante07}.

RPS can be a significant contributor to metal injection in the ICM.
From the theoretical side, the ISM of galaxies will be stripped if the ICM density
is as low as 5 $\times 10^{-4}$ atoms cm$^{-3}$ \citep{GG72}. This rough estimate
(other factors, such as the direction along which the galaxy is traveling through the ICM, may affect
this number) suggests that RPS could be effective at large spatial scales in most galaxy clusters \citep{Tonnesen08}.

Support for RPS effectiveness comes from recent observations as well.
\citet{Haynes84}, \citet{Bravo00} and \citet{Sun07} noticed that the HI 
emission is sharply truncated in spirals of 
cluster of galaxies, indicating significant RPS. \citet{Bruggen08} found that 
about one quarter of galaxies in massive clusters are subject to strong RPS that are 
likely to cause an expedient loss of all gas. Moreover, there are recent observations indicating that
RPS is effective in removing the galaxy gas even in low density environments, such as groups and in the 
outskirts of clusters \citep{Solanes01,Kantharia05,Levy07,Kantharia08}, where this mechanism should not, in principle, 
play a significant role.
Using hydrodynamical simulations to study RPS, \citet{McCarthy08} found that $\sim$ 70\% of the galactic halo
is stripped, for typical structural and orbital parameters. 

\begin{figure}
\centering
\includegraphics[width=0.37\textwidth,angle=90]{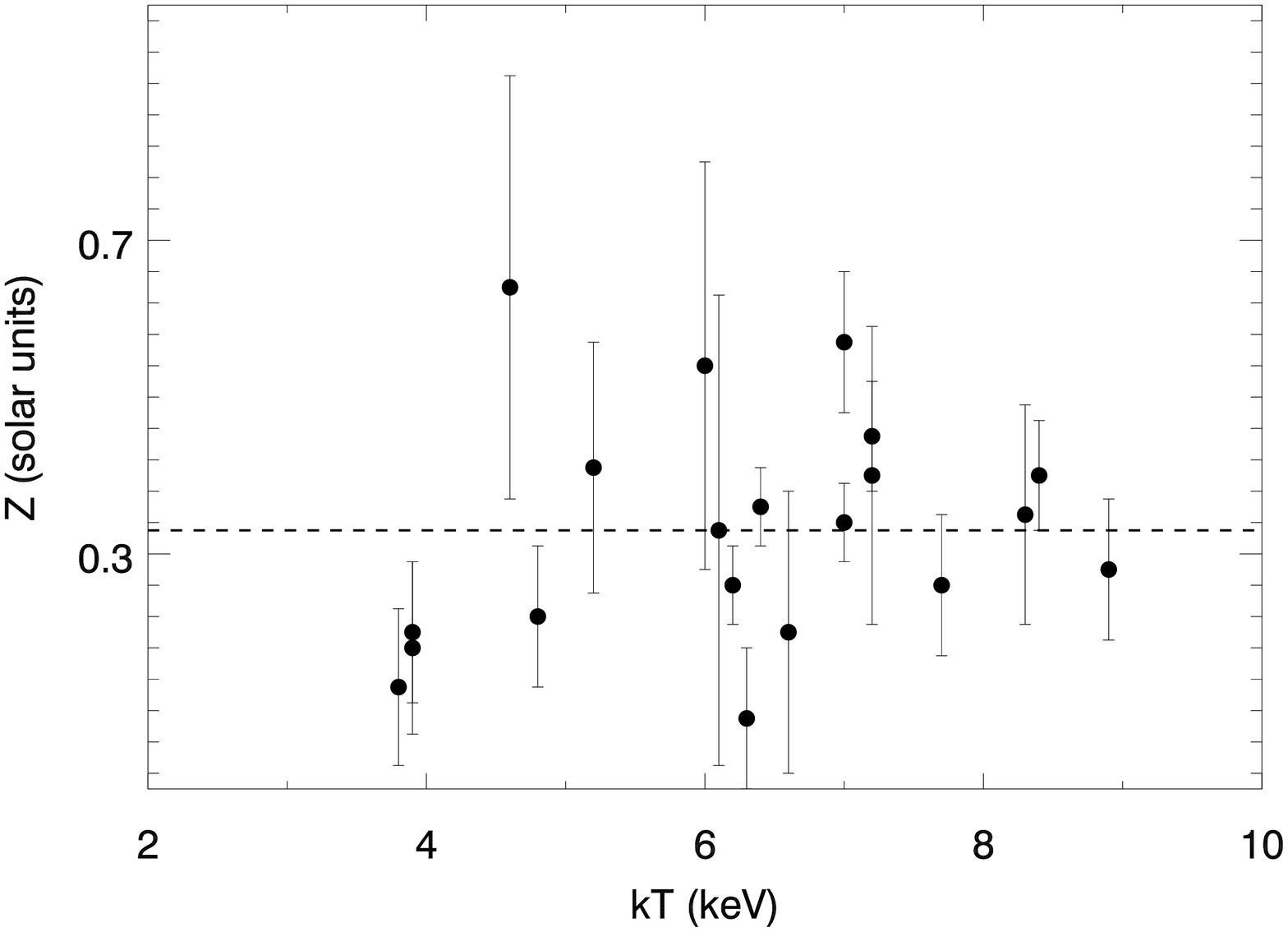}
\caption{\small Metallicity as a function of temperature for the clusters analyzed in this work. The dashed line represents is the mean value of the iron abundance in this sample.}
\label{MfekT}
\end{figure}

If RPS is important and has a dependence on cluster mass we should
see a trend between gas temperature and metallicity, since the higher the mass 
(or the temperature) the more
metals are stripped from galaxies and injected into the ICM. 
However, we do not see this trend (Fig.~\ref{MfekT}) within the temperature range of 
our sample ($2< kT$~(keV)~$< 10$).
Therefore, either RPS is negligible, which contradicts the theoretical expectations and the 
observations described in the previous paragraphs, or RPS is equally significant for all clusters,
once the threshold ICM density is reached. Since RPS is not biased by galaxy morphology this
would be consistent with the lack of a correlation between iron mass and galaxy morphology.

Chemical evolution models have been invoked to analyze the iron metallicity, suggesting
that the number of SN Ia that ever occurred relative to SN II is $N_{\rm Ia}/N_{\rm II}=0.12$,
with more than 50\% of the $^{56} \rm Fe$ mass coming from SN Ia \citep{Yoshi96,Nomoto97}. 
The iron yield ($p$) is given by \citep{Arnaud92}:
\begin{equation}
[M_{\rm Fe}]_{\rm ICM} + [M_{\rm Fe}]_{\star} \le p ~ M_{\star},
\end{equation}
where the two terms in the left side are the total iron mass in the ICM, $[M_{\rm Fe}]_{\rm ICM}$, and in stars, 
$[M_{\rm Fe}]_{\star}$, and $M_{\star}$ is the total stellar mass.
Since the ratio $([M_{\rm Fe}]_{\rm ICM} + [M_{\rm Fe}]_{\star})/M_{\star}$ cannot exceed the iron yield, if we assume 
that all the iron ever produced goes to the ICM, we have,
\begin{equation}
\frac{[M_{\rm Fe}]_{\rm ICM}}{M_{\star}} \le p,
\end{equation}
that is, $M_{\rm Fe}/M_{\star}$ gives an upper limit for the iron yield, since 
some recycling must have occurred. 

The total mass in stars can be computed by multiplying the total luminosity 
($L_{\rm red}+L_{\rm blue}$) by an apropriate mass-to-light ratio. The late-type and early-type 
mass-to-light ratios were estimated from \citet{kauff03} and converted to the $r$ band according 
to \citet{Fukugita95}, and are $M/L = 3.27 M_{\odot}/L_{\odot}$ for a red population and 
$M/L = 1.64~ M_{\odot}/L_{\odot}$
for a blue population. This procedure was explained in details in \citet{Lagana08}.

For our sample the mean iron-to-stellar mass ratio is 
$[M_{\rm Fe}]_{\rm ICM}/M_{\star} = (6.0 \pm 0.5)\times 10^{-3}$. \citet{DFJ90} calculated the
amount of iron that type II driven winds can inject in the ICM.  These authors obtained 
$[M_{\rm Fe}]_{\rm ICM}/M_{\star}$ ranges from $0.65 \times 10^{-3}$ to $2.5 \times 10^{-3}$, 
depending on the IMF assumption. 
Ours result show much higher values
suggesting that 
more than 50\% of the iron present in the ICM is probably produced 
in SN Ia. 
Most of the iron produced by SN Ia is preferentially 
injected into the ICM by RPS.
Since RPS is equally efficient within the temperature range of our sample,
the suggestion that more than half of the iron mass is produced in type Ia SN 
corroborate the result of a lack of
trend between the metallicity and the ICM temperature.

This last result is consistent with models with models in which metals might have been accumulated 
in the ICM in two phases \citep{White91,Elbaz95,Matteucci95,Dupke00}: an initial phase 
of SN II activity, responsible for part of the iron found in the ICM, followed by a 
secondary phase, associated to the BCG formation, 
contributing with more than 50\% of the ICM iron, where most of the metals were produced by SN Ia 
and injected in the ICM by RPS.

\section{Conclusions}
\label{conc}

We have carried out a detailed analysis of the iron mass in the ICM and its correlation with 
optical properties for 20 galaxy clusters previously studied by \citet{M08} and available in the SDSS. 
Our main results are:
\begin{itemize}

\item{We could not confirm a previous correlation between the ICM iron mass and the total luminosity of the 
red population, found by \citet{Arnaud92}. Since our results indicate that the BCGs seem to play a major role in the 
ICM iron enrichment,  we suggest that the trend found by \citet{Arnaud92} is biased by the BCGs as these authors did not
exclude them from the `E+S0' population. 
As the BCGs alone cannot produce the observed metallicity within $r_{500}$, we 
explain the correlation between the iron mass and the BCG luminosities 
through a scenario in which a mechanism simultaneously enhances the luminosity of
the BCG and the iron mass in the ICM. We suggest RPS with tidal disruption
near the cluster center as a possible mechanism, the importance of which is supported by recent hydrodynamical
simulations \citep[mechanism III of][]{Murante07}.}

\item{The lack of trend between the iron metallicity and the cluster temperature indicates that RPS 
is equally efficient in all clusters within the range 2 $< \rm kT (keV) <$ 10. 
This is in agreement with current observational and theoretical \citep{Tonnesen08} studies, which
suggest that RPS is more common in low 
density environments than previously thought. Thus, RPS is a significant 
mechanism for transferring iron from galaxies to the intra-cluster gas.}

\item{The comparison of our results with predictions of chemical evolution models 
suggests that more than 50\% of the iron has come from type Ia SNe. 
Since most of the iron produced by SN Ia is preferentially 
injected into the ICM by RPS, the iron yield corroborates the efficiency of RPS
within the temperature range of this sample analysed here.
There have been observational evidence, supported by hydrodynamical simulations, 
that the RPS mechanism is contributing to the gas removal from galaxies 
that merged to form the BCGs \citep{Murante07}.}

\item{We suggest that the complex history of galaxy populations in clusters, from galaxy infall
followed by the action of severe environmental effects, leads
to galaxy morphological (and colour) evolution and, at the same time, to a progressive 
enrichment of the ICM, diluting the role of any single population.
Clearly, larger samples are needed to verify if these correlations are actually real.}

\end{itemize}

\section{acknowledgments}
The authors thank B. J. Maughan for making available the best fit parameters of the density profiles
and the anonymous referee for constructive suggestions.
The authors also acknowledge financial support from the Brazilian agencies FAPESP, CNPq
and CAPES (grants: 03/10345-3 and BEX1468/05-7), as well as the Brazilian-French collaboration 
CAPES/Cofecub (444/04).  R. Dupke acknowledges partial support from NASA (grants: GO5-6139X,
NNX06AG23G, NNX07AH55G and NNX07AQ76G). We also wish to thank the team of the
Sloan Digital Sky Survey (SDSS) for their dedication to a project 
which has made the present work possible.

\bibliographystyle{mn2e}
\bibliography{Lagana08_final}

\end{document}